\documentclass[journal,transmag]{IEEEtran}

\usepackage{amsmath}
\usepackage{graphicx}
\usepackage{caption}
\usepackage{subcaption}
\usepackage[list=false]{subcaption} % 导言区添加
\usepackage{subcaption}
\usepackage{array}
\usepackage{amssymb}
\usepackage{placeins}
\usepackage{xcolor}
\usepackage{multirow}
\usepackage{hyperref}
\usepackage{booktabs} % for nicer table rules
\usepackage{makecell}   % Allows line breaks in table cells

\usepackage[margin=1in]{geometry}
\usepackage{multirow}
\begin{document}
\title{Dynamic hysteresis model of grain-oriented ferromagnetic material using neural operators}

\author{
\IEEEauthorblockN{
Ziqing Guo\IEEEauthorrefmark{1}\IEEEauthorrefmark{2},
Binh H. Nguyen\IEEEauthorrefmark{3},
Hamed Hamzehbahmani\IEEEauthorrefmark{4},
Ruth V. Sabariego\IEEEauthorrefmark{1}\IEEEauthorrefmark{2}
}
\IEEEauthorblockA{
\IEEEauthorrefmark{1}Department of Electrical Engineering, KU Leuven, 3000 Leuven, Belgium \\
\IEEEauthorrefmark{2}EnergyVille, 3600 Genk, Belgium \\
\IEEEauthorrefmark{3}Institute of Engineering Mathematics, University Duisburg-Essen, 45141 Essen, Germany \\
\IEEEauthorrefmark{4}Dept. of Engineering, Durham University, Durham DH1 3LE, United Kingdom
}
\thanks{Corresponding author: Z. Guo (email: ziqing.guo@kuleuven.be).}
}
% The paper headers
% \markboth{Journal of \LaTeX\ Class Files,~Vol.~14, No.~8, November~2024}%
% {Shell \MakeLowercase{\textit{et al.}}: Bare Demo of IEEEtran.cls for IEEE Transactions on Magnetics Journals}

\IEEEtitleabstractindextext{%
\begin{abstract}
%%%%%%%%%%%%%%%%%%%%%%%%%
Accurately capturing the behavior of grain-oriented (GO) ferromagnetic materials is crucial for modeling the electromagnetic devices. In this paper,  neural operator models, including Fourier neural operator (FNO), U-net combined FNO (U-FNO) and Deep operator network (DeepONet) are used to approximate the dynamic hysteresis models of GO steel. 
Furthermore, two types of data augmentation strategies including cyclic rolling augmentation and Gaussian data augmentation (GDA) are implemented to enhance the learning ability of models. 
With the inclusion of these augmentation techniques, the optimized models account for not only the peak values of the magnetic flux density but also the effects of different frequencies and phase shifts.
The accuracy of all models is assessed using the L2-norm of the test data and the mean relative error (MRE) of calculated core losses. Each model performs well in different scenarios, but FNO consistently achieves the best performance across all cases.
\end{abstract}
}
% make the title area
\maketitle
\IEEEdisplaynontitleabstractindextext
\IEEEpeerreviewmaketitle
\section{Introduction}
\IEEEPARstart{C}{omprehensive} modeling of electromagnetic devices with laminated cores is essential for their analysis and optimization \cite{Ferr_core_loss_Roshen_1991, chen_dynamic_hysteresis_model_2022}.
Phenomena such as saturation, hysteresis and eddy current greatly increase the difficulty of core modeling and analysis \cite{zirka_dynamic_JA_2015}. 
For simplicity, most of the models treat permeability as constant and obtain the core losses in post-processing \cite{Florent_liege_NN_2024, cardelli_comparison_hysteresis_models_2023}.
% While, in certain practical applications, 
These simplification do not justify in some practical applications where the
core losses are not only significant but also dominate other types of power losses \cite{corti_dynamic_core_losses_2020}. Accurate, efficient and straightforward electromagnetic modeling that accounts for hysteresis and induced eddy-current losses in magnetic cores is crucial \cite{cardelli_comparison_hysteresis_models_2023}.\\
From the modeling perspective, the mathematical hysteresis models such as Jiles-Atherton model \cite{jiles1986theory} and Preisach \cite{Mayergoyz_dynamic_preisach_1988} model are widely recognized and used by engineers and physicists in the past several decades. They are usually coupled to finite element method (FEM) for the computation of the dynamic losses in magnetic cores through solving the Maxwell equations or the simplified 1D diffusion equation (for a thin homogeneous ferromagnetic strip) \cite{cardelli_comparison_hysteresis_models_2023}.
Unfortunately, at power frequencies, the discrepancy between measured and calculated losses is often negeligible for non-oriented (NO) magnetic cores. However, for grain-oriented (GO) materials, which constitute over $90\%$ of transformer cores, this difference can reach approximately $40\%$ \cite{zirka_static_dynamic_hysteresis_model_2011}.
Numerous attempts to build a dynamic model of GO steel strip goes back to 1950s and the milestone study was proposed by Bertotti in 1998 \cite{Bertotti_loss_separation_1988}.
%%%%%%%%%%%%%%%%%%%%%%%%%%%%%%%%%%%%%%%%%%%%%%%%%%%%%%%%%%%%
In Bertotti's theory, the total power losses can be separated into 3 parts: the hysteresis losses, classical eddy current losses and the excess losses, which provides an insight into power losses mechanism of soft magnetic materials. During the last decades, statistical losses separation theory of Bertotti has been broadly developed and implemented as a theoretical background in dynamic modeling of soft magnetic materials. Zirka et al. \cite{zirka_dynamic_JA_2015} showed that power losses separation theory can be mathematically interpreted to magnetic field separation, which implies that the magnetic field $H$ can be separated into hysteresis field, classical eddy current field and excess field. This approach has been extensively used in dynamic modeling of GO electrical steels with high accuracy \cite{hamed_GO_model_2021}. 
%%%%%%%%%%%%%%%%%%%%%%%%%%%%%%%
However, as the basis of dynamic hysteresis loops (DHLs), hysteresis field is not easily and accurately to be established. It can be represented by the static hysteresis loop (SHL) or quasi-static hysteresis loop ($f$ ranging from 0.004\,Hz to 5 \,Hz \cite{zirka_static_dynamic_hysteresis_model_2011, du_dynamic_JA_losses_high_freq_2014}). One practical technique to measure the SHL is to magnetize the material at a very low magnetizing rate to eliminate the dynamic field. However, first, measuring quasi-static hysteresis loop at low and very low frequencies may not be always feasible due to the requirement of special equipment or bandwidth limitation of the test setup \cite{hamed_GO_model_2021}. Second, the formulation of the dynamic terms especially of the excess field involves multiple steps, requiring different forms of equations and parameters estimation based on the frequency and peak values of the collected data \cite{hamed_GO_model_2021}. As a result, developing a dynamic hysteresis model with both high accuracy and strong generalization ability is crucial.\\
Recently, with the development of data-driven methods and machine learning, numerous of data-driven hysteresis models are proposed. There are several types of neural networks (NNs) used for modeling the hysteresis including the backpropagation to model the DHL with different frequencies \cite{tian_backpropogation_NN_2021}, the Preisach model with recurrent neural network (RNN) approximating the density functions \cite{grech_RNN_Preisach_2020} and long short term memory (LSTM) network with convolutional neural network (CNN) extracting high-dimensional data features reflecting the hysteresis characteristics of the loop \cite{ding_LSTM_CNN_2024}. 
However, most of the models are focusing on frequencies as 500 Hz. In \cite{Magnet_Haoran_2023}, encoder-decoder is used for the building the hysteresis model with the frequency ranging from 50 to 500 kHz, which shows the great ability of NN for modeling hysteresis with high frequency for N87 Ferrite rather than GO steel.\\
A new type of neural operator based on the universal operator approximation theory is proposed in \cite{chen_universal_app_theory_1995}. This neural operator is very powerful for nonlinearity modeling as it learns the operator itself instead of approximating the solution of one specific equation \cite{lulu_Deeponet_2021}. 
There are mainly 2 types of neural operator, deep operator network (DeepONet) \cite{lulu_deeponet_2019} and Fourier neural operator (FNO) \cite{zongyi_FNO_2020}, which have been successfully applied in fluid dynamics \cite{Lyu_fluid_flow_NO_2023}, piezoelectric dynamics \cite{TU_Delft_NO_2024} and heat transfer \cite{yuan_heat_transfer_NO_2025} but rarely in electromagnetism. Very recently, the static hysteresis model based on neural operator was proposed, which validated the feasibility and the generalization ability of neural operator for static hysteresis model \cite{chandra_no_magnetic_2024}.
In this paper, FNO, DeepONet and U-FNO, a type of expanded FNO, are employed to build the dynamic hysteresis model for GO steel. The accuracy of the models are compared based on the errors between the predictions and the reference values of measured test data.
The structure of the paper is as follows: In section \ref{sec:energy_loss_separation}, the energy losses separation theory and the feasibility of neural operator for DHLs modeling are illustrated and analyzed; In section \ref{sec:Neural operator models}, the principle and the structure of FNO, U-FNO and DeepONet are explained in detail. Then, the experiments and the results are presented in section \ref{sec:experiment_and_results}, including the initial results and the optimized results with data augmentation strategies. Some conclusions are drawn in the last section.
\section{Energy loss separation theory}
\label{sec:energy_loss_separation}
As illustrated in energy losses separation theory \cite{Bertotti_loss_separation_1988}, there are three components, $W_{hys}$, $W_{eddy}$ and $W_{exc}$, where $W_{hys}$ is the hysteresis energy losses, $W_{eddy}$ is the classical eddy current losses, and $W_{exc}$ is the excess energy losses, 
which leads to the magnetic field $H$ with hysteresis field $H_{hys}$, eddy-current field $H_{eddy}$ and excess field $H_{exc}$ respectively as \cite{hamed_GO_model_2021}
\begin{equation}
\label{equ:H_3parts}
    H(t) = H_{hys}(t) + H_{eddy}(t) + H_{exc}(t)
\end{equation}
These three losses components are active simultaneously in any ferromagnetic material subject to a time-varying magnetic field \cite{hamed_GO_model_2021}. In GO steels, which have relatively large grain sizes, excess losses contribute significantly to the total energy losses. This makes accurate computation of losses more challenging due to the increased uncertainty associated with excess losses \cite{zirka_static_dynamic_hysteresis_model_2011}.\\
% excess energy losses counts for a significant amount of energy losses \cite{zirka_static_dynamic_hysteresis_model_2011}.\\
Using the dynamic models of the classical eddy current and excess fields, combined with the magnetic constitutive law, Eq. \eqref{equ:H_3parts} can be expressed as the thin sheet model for ferromagnetic materials \cite{zirka_dynamic_JA_2015},
% Using the dynamic models of the classical eddy current and excess fields, Eq. \eqref{equ:H_3parts} can be re-expressed as the thin sheet model (TSM) for ferromagnetic material \cite{zirka_dynamic_JA_2015}:
\begin{equation}
\label{equ:H_3parts_equations}
    % H(t) = H_{\text{hys}}(B(t)) + \frac{\frac{m^2 B}{dt^2}}{12 \rho} + g(B) \, \delta \, \left| \frac{dB}{dt} \right|^{\alpha}
    H(t) = H_{hys}(B(t)) + \frac{m^2}{12\rho}\frac{dB}{dt} + g(B)\delta \left |\frac{dB}{dt}\right |^\alpha ,
\end{equation}
where $m$ is the lamination thickness, $\rho$ is the material resistivity, and directional parameter $\delta = +1$ for $(dB/dt)>0$ and  $\delta = -1$ for $(dB/dt)<0$, the exponential coefficient $\alpha$ designates the frequency dependence of the excess field component, and $g(B)$, in general, is a polynomial function of the flux density $B$ to control shape of the modeled hysteresis loop. The first term in Eq. \eqref{equ:H_3parts_equations}, $H_{hys}(B(t))$ is the static hysteresis model, which is independent from frequency but depends on the instantaneous values of the magnetic field $B(t)$ for each magnetizing frequency, and can be determined from Jiles-Atherton model \cite{Marion_JA_model_parameters_2008}.
On the one hand, Eq. \eqref{equ:H_3parts_equations} describes a complex hysteresis behavior in GO ferromagnetic materials, where the magnetic field depends on the magnetic flux both implicitly and explicitly, such that identifying the model parameters in a non-trivial task \cite{zguo_PINN_2025}. On the other hand, Eq. \eqref{equ:H_3parts_equations} can be interpreted as a nonlinear mapping from the magnetic flux to the magnetic field, which can be approximated as a neural operator.
\section{Neural operator models}
\label{sec:Neural operator models}
\begin{figure*}[!htbp]
    \centering
    
    % 第一个子图
    \begin{subfigure}[t]{0.80\textwidth}
        \centering
        \includegraphics[width=\textwidth]{ 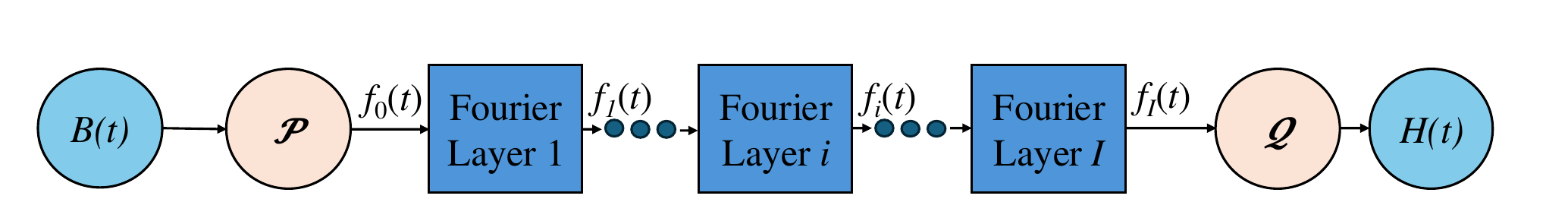}
        \phantomcaption  % 关键修改：隐藏 (a) 但保留标签锚点
        \label{fig:FNO_general_structure}
    \end{subfigure}
    \par\medskip
    \begin{minipage}{\textwidth}
        % \centering
        \small
        \textbf{(a)} FNO structure. With $B(t)$ as the input, $\mathcal{P}$ as a linear NN for lifting $H(t)$, $I$ Fourier layers with $f_i(t)$ as input and $f_{i+1}(t)$ as output, $\mathcal{Q}$ as another linear NN for projecting the transformed data into output $H(t)$.
    \end{minipage}

    \vspace{0.5cm}

    % 第二个子图
    \begin{subfigure}[t]{0.90\textwidth}
        \centering
        \includegraphics[width=\textwidth]{ 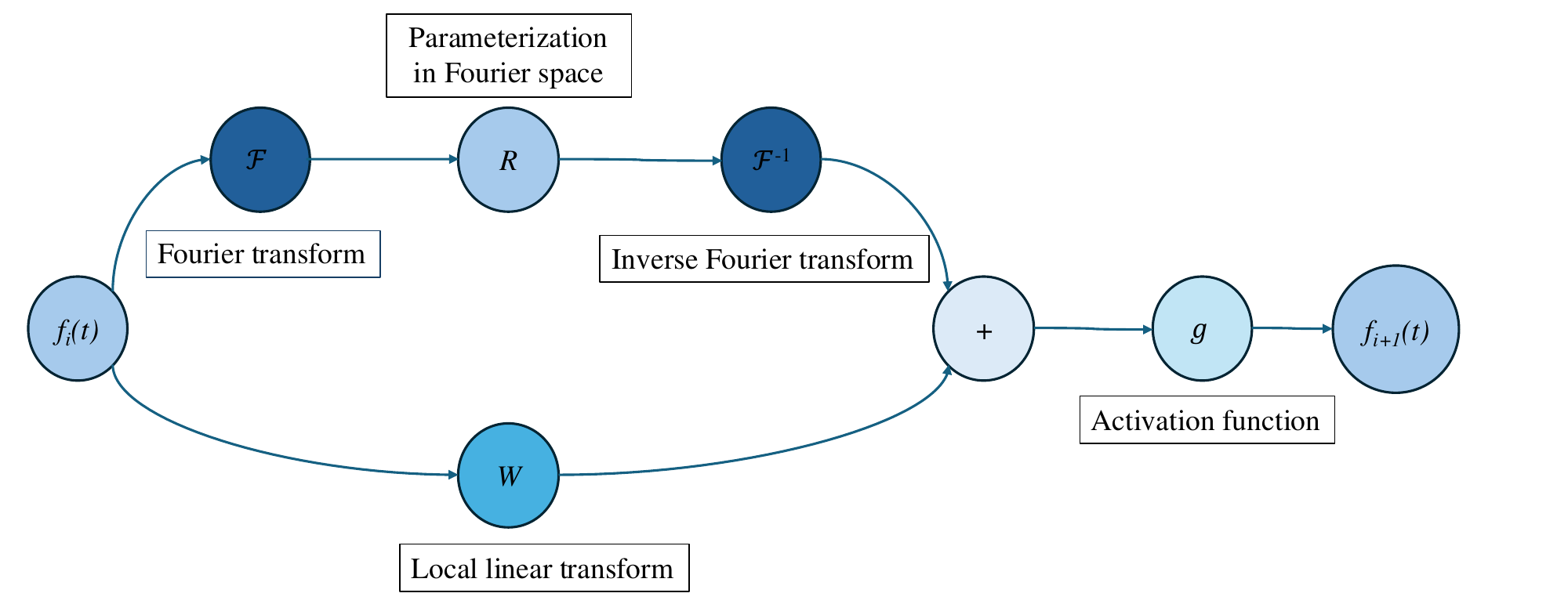}
        \phantomcaption  % 隐藏 (b) 但保留标签锚点
        \label{fig:FNO_Fourier_structure}
    \end{subfigure}
    \par\medskip
    \begin{minipage}{\textwidth}
        % \centering
        \small
        \textbf{(b)} FNO layer. With $\mathcal{F}$ as the Fourier transform, $R$ as the parameterization in Fourier space, $\mathcal{F}^{-1}$ as the inverse Fourier transform, $g$ as the nonlinear activation function, the residual connection by adding linear transformation $W$.
    \end{minipage}

    \caption{FNO structure and unit.}
    \label{fig:UFNO_structure}
\end{figure*}

\begin{figure*}[!htbp]
    \centering
    
    % 第一个子图
    \begin{subfigure}[t]{0.6\textwidth}
        \centering
        \includegraphics[width=\textwidth]{ 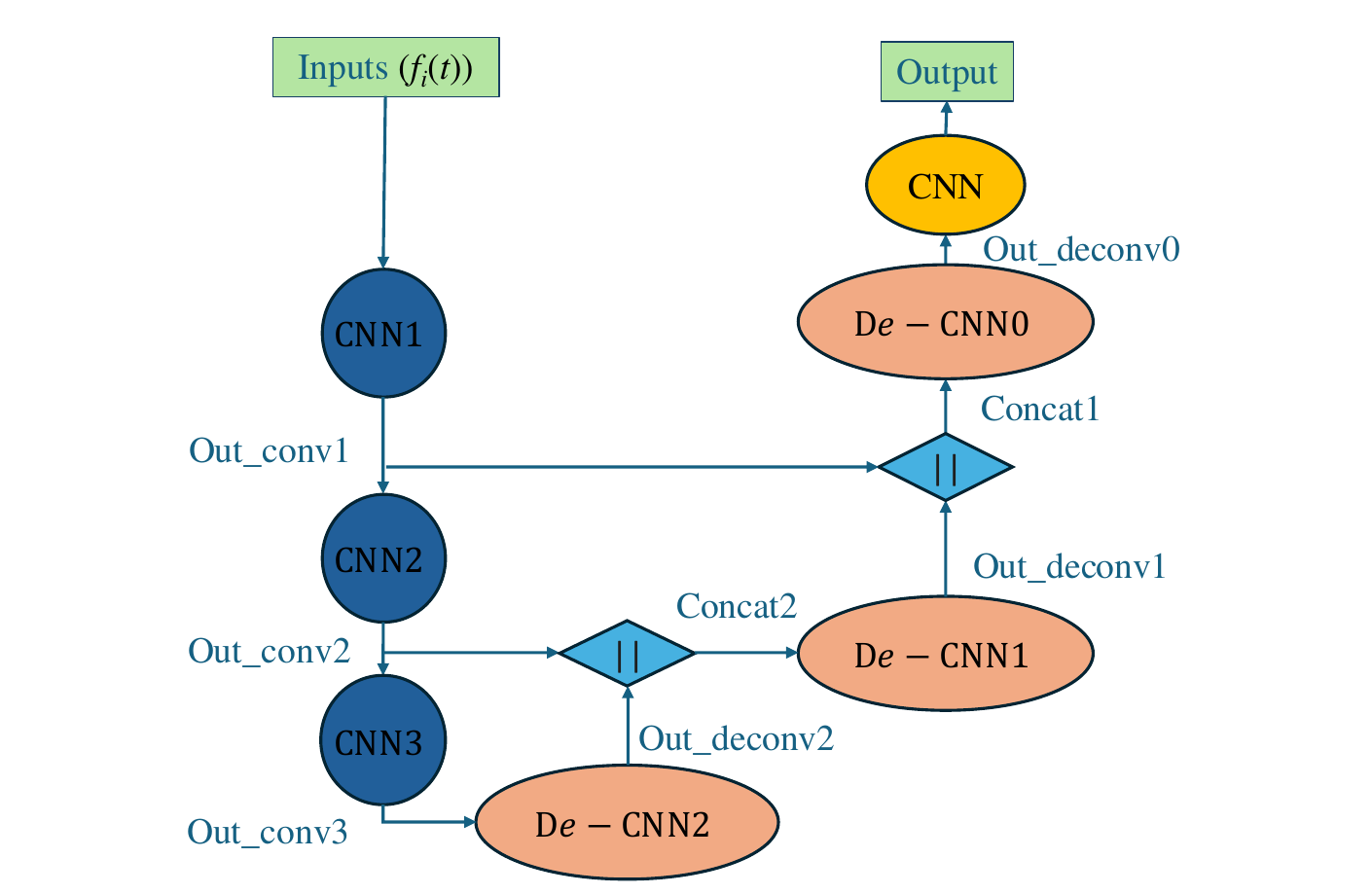}
        \phantomcaption  % 关键修改：隐藏 (a) 但保留标签锚点
        \label{fig:U_net_structure}
    \end{subfigure}
    \par\medskip
    \begin{minipage}{\textwidth}
        % \centering
        \small
        \textbf{(a)} U-net structure. With 3 convolutional NNs (CNN1, CNN2, CNN3) and 3 deconvolutional NNs (De-CNN0, De-CNN1, De-CNN2). $||$ representing concatenation.
    \end{minipage}

    \vspace{0.5cm}

    % 第二个子图
    \begin{subfigure}[t]{0.90\textwidth}
        \centering
        \includegraphics[width=\textwidth]{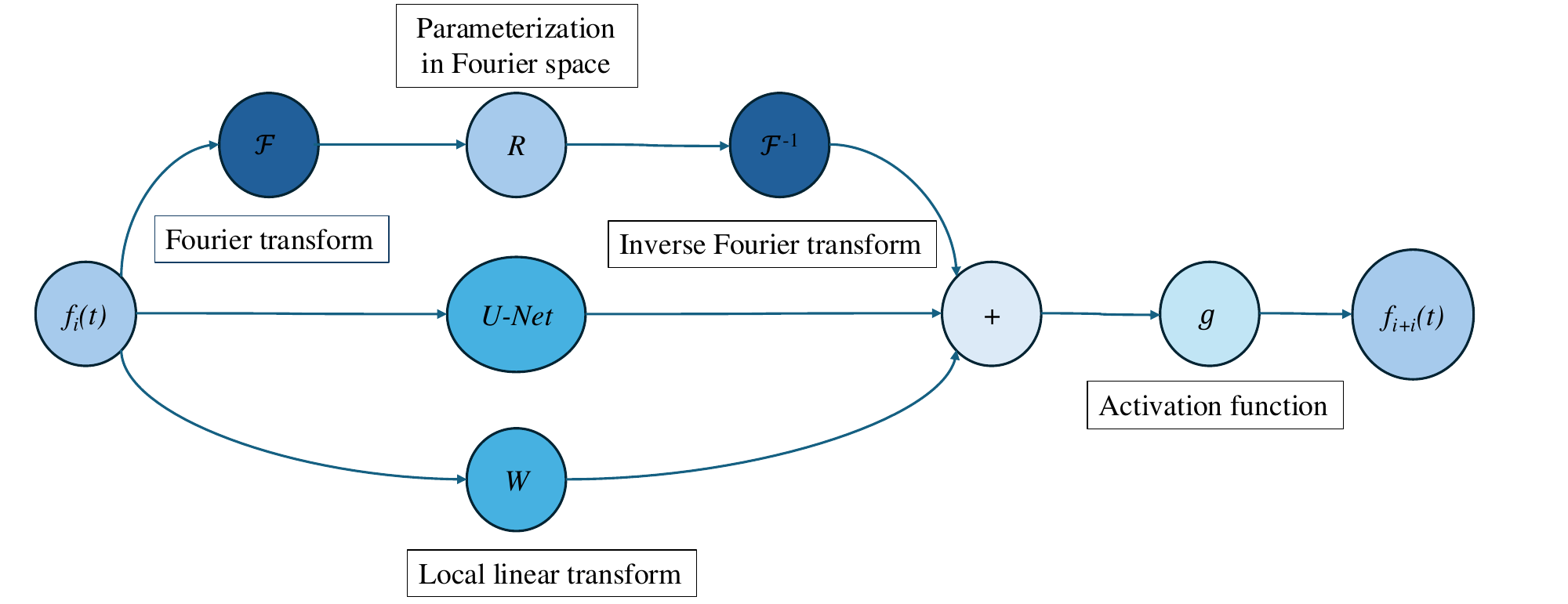}
        \phantomcaption  % 隐藏 (b) 但保留标签锚点
        \label{fig:U-FNO layer}
    \end{subfigure}
    \par\medskip
    \begin{minipage}{\textwidth}
        % \centering
        \small
        \textbf{(b)} U-FNO layer. With U-net as in Fig.~\ref{fig:U_net_structure}. $\mathcal{F}$: Fourier transform; $R$: parameterization in Fourier space; $\mathcal{F}^{-1}$: inverse Fourier transform; $g$: nonlinear activation; residual connection via linear transformation $W$.
    \end{minipage}
    \caption{U-net structure and U-FNO unit. }
    \label{fig:UFNO_structure}
\end{figure*}
The neural operator is proposed based on the universal operator approximation theorem \cite{chen_universal_app_theory_1995}, it learns a mapping between two infinite dimensional spaces from a finite collection of observed input-output pairs. Let \( D \subset \mathbb{R}^d \) be a bounded, open set and \( \mathcal{A} = \mathcal{A}(D; \mathbb{R}^{d_a}) \) and \( \mathcal{U} = \mathcal{U}(D; \mathbb{R}^{d_u}) \) be separable Banach spaces of functions taking values in \( \mathbb{R}^{d_a} \) and \( \mathbb{R}^{d_u} \) respectively. Furthermore, let \( G^\dagger : \mathcal{A} \to \mathcal{U} \) be a typically nonlinear map. The objective of neural operator is to build an approximation of \( G^\dagger \) by constructing a parametric map\cite{kovachki_zongyi_FNO_2023}, it reads:
\begin{equation}
\label{eq:G+}
    G^\dagger: \mathcal{A} \times \Theta \to \mathcal{U} \quad\,,
\end{equation}
where $\Theta$ is the finite-dimensional parameter space.
The cost function is defined as the error between the approximation of the built operator and the real operator, which decreases during training to choose the optimal parameters $\theta^\dagger \in \Theta$ of neural operator network.
\subsection{Fourier neural operator (FNO)}
\label{sec:FNO}
The FNO \cite{zongyi_FNO_2020} is designed to parameterize the integral kernel directly in Fourier space, learning the Fourier coefficients from data. FNO discretizes both the input ($B(t)$) and output ($H(t)$) on an equispaced mesh and uses the same mesh for discretization. As shown in Fig. \ref{fig:FNO_general_structure}, FNO first lifts the input $B(t)$ to a higher-dimensional representation $f_0(t)$ using a shallow fully connected layer ($\mathcal{P}$). Then, $I$ fourier layers are applied iteratively. Each Fourier layer performs as in Fig. \ref{fig:FNO_Fourier_structure}:\\
\begin{itemize}
    \item The input $f(t)$ is first transformed into the frequency domain with selected modes by Fourier transform ($\mathcal{F}$);
    \item Parameterization in Fourier space: trainable parameters (the weights and the bias of NN) are applied in the frequency space, which is modified during training by decreasing the loss function;
    \item The modified frequency representation is converted back to time space by the inverse Fourier transform ($\mathcal{F}^{-1}$);
    \item The residual connection \cite{kaiming_residual_connection_2016} is used by applying a parallel linear transformation to the physical space representation for further refinement. The output from this linear bias term are then element-wise added with the output of inverse Fourier transform. Nonlinear activation is used to introduce nonlinearity and improve learning capacity.    
    \item Finally, after all Fourier layers, another linear layer ($\mathcal{Q}$) projects the transformed data to the output $H(t)$. 
\end{itemize}
This architecture allows FNO to efficiently handle structured grids, making it well-suited for learning operators. 
% It is important to note that Fourier layers are discretization-invariant, meaning they can learn from and evaluate functions independently of the discretization.
\subsection{U-net combined FNO (U-FNO)}
\label{sec:U_FNO}
U-net is a neural network architecture initially designed for image segmentation \cite{siddique_U_NET_2021}. The basic structure of a U-net architecture consists of encoder and decoder. The encoder, also called contracting path, is similar to a regular convolution network and provides classification information. 
The novelty of U-net comes in the decoder, also called the expansive path, in which each stage upsamples the feature map using up-convolution. Then, the feature map from the corresponding layer in the contracting path is cropped and concatenated onto the upsampled feature map \cite{siddique_U_NET_2021}. The resulting network is almost symmetrical, giving it a U-like shape as shown in Fig \ref{fig:U_net_structure}.
As shown in Fig. \ref{fig:U-FNO layer}, this U-net is added into the FNO-layer, parallel to the linear bias term and the Fourier layer to build the U-FNO layer \cite{zongyi_U_FNO_2022}. This expansion can process local convolution to enrich the representation ability of NN for local features \cite{siddique_U_NET_2021}. The general structure of U-FNO is similar to FNO as Fig. \ref{fig:FNO_general_structure}, but replacing several FNO layers by the U-FNO layers as Fig. \ref{fig:U-FNO layer}. The number of FNO layer and U-FNO layer are hyperparameters that can be optimized for the specific problem.\\
\subsection{Deep Operator Network (DeepONet)}
\label{sec:DeepONet}
\begin{figure}[!h]
    \centering
    \includegraphics[width=0.8\linewidth]{ 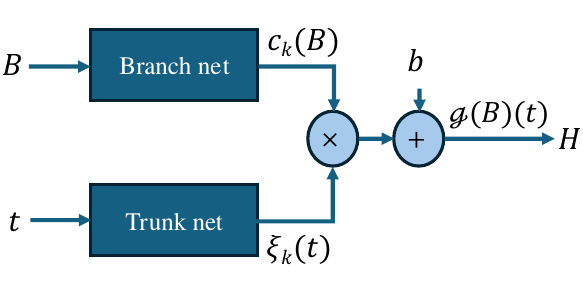}
    \caption{DeepONet structure. Branch net with $B$ and $c_k(B)$ as input and output, trunk net with the coordinates $t$ and 
$\xi_k{t}$ as the input and output, $b$ as the bias, $\mathcal{G}(B)(t)$ as the approximated output $H$.}
    \label{fig:DeepONet_structure}
\end{figure}
DeepONet is first proposed in \cite{lulu_deeponet_2019}. As shown in Fig. \ref{fig:DeepONet_structure}, there are 2 sub-networks in DeepONet: the branch net taking the discretized function $B$ as input, and trunk net with the coordinates $t$ as the inputs. Finally, the output of the two sub-networks are multiplied to generate the final output with adding bias $b$, which can be expressed as
\begin{equation}
\label{equ: deeponet_output}
    \mathcal{G}(B)(t) = \sum_{k=1}^p c_k(B) \xi_k(t) + b\,,
\end{equation}
where $b \in \mathbb{R}$ is bias. $\{c_1, c_2, \ldots, c_p\}$ are the $p$ outputs of the branch net, and $\{\xi_1, \xi_2, \ldots, \xi_p\}$ are the $p$ outputs of the trunk net. As $t$ is usually low dimensional, a standard FNN is commonly used as the trunk net. The choice of the branch net depends on the structure of the input function $B$, it can be chosen as any type of NN like FNN, ResNet, CNN, RNN, or a graph neural network (GNN), etc.\\
\section{Experiments and results}
\label{sec:experiment_and_results}
\subsection{Experiment setup and data information}
\label{sec:data_information}
\begin{figure}[!h]
    \centering
    \includegraphics[width=1.0\linewidth]{ 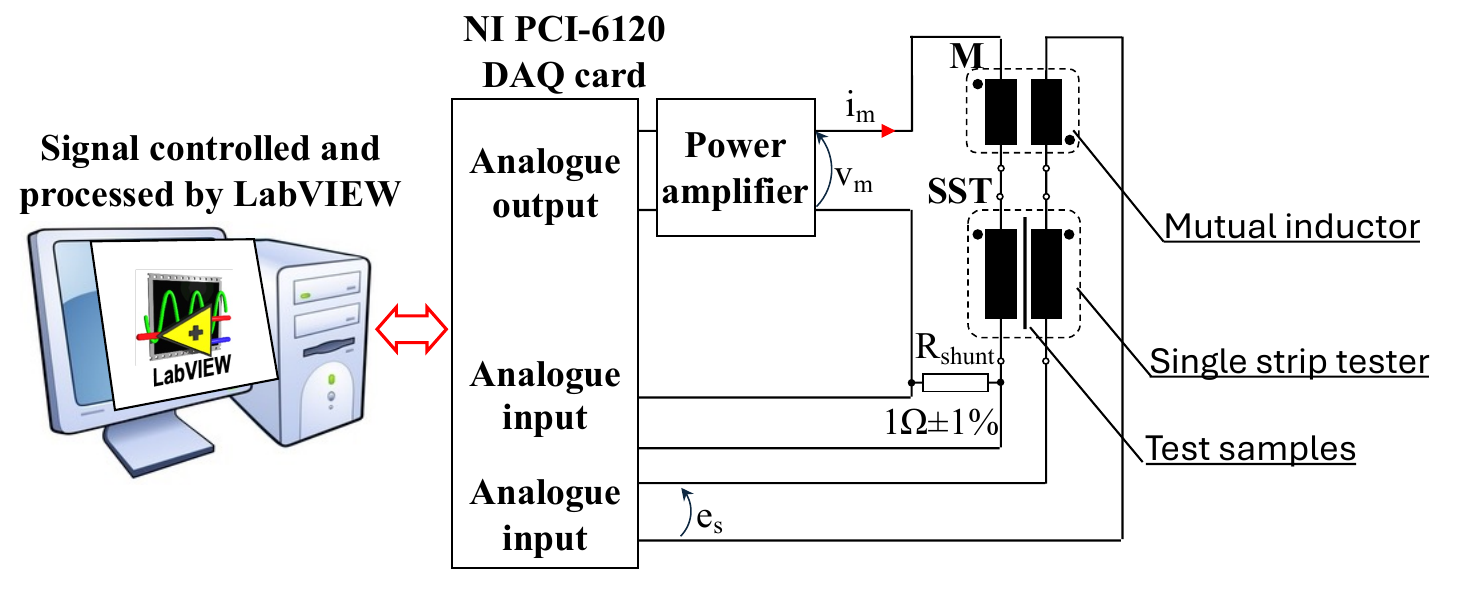}
    \caption{The measuring system. It comprises a personal computer, a NI PCI‐6120 data acquisition (DAQ) card, an audio power amplifier, and an air‐flux compensated single strip tester (SST).}
    \label{fig:test_setup}
\end{figure}

% A 0.3\,mm thick standard grades of M105-30P GO with $3\%$ SiFe and the resistivity as $\rho = 0.461 \mu \Omega \text{m}$ was used for experiment. 
Fig. \ref{fig:test_setup} is the schematic diagram of the test setup, which consists of a single strip test (SST) controlled by a LabVIEW-based system using a NI PCI-6120 data acquisition (DAQ) card. Standard test samples of M105-30P GO, 0.3\,mm thick with $3\%$ SiFe and resistivity $\rho = 0.461 \mu \Omega \text{m}$ were used for the experiment.
An SST was used to magnetize the test samples under controlled sinusoidal inductions.
The DHLs of the GO at 9 frequencies 5, 10, 25, 50, 100, 200, 400, 800 and 1000\,Hz and magnetic flux density amplitudes 1.0, 1.3, 1.5, 1.7\,T for each frequency were collected. So, there were totally $9\times4 = 36$ DHLs collected from experiments, i.e. one period in steady state with 500 samples each. Note that all $B$ and $H$ are normalized to improve the training performance of the model. Training was carried out on CPU of an Apple M1 Max laptop with an ARM-based architecture. 
\subsection{Initial results of the neural operator}
\subsubsection{FNO}
\label{sec:results of FNO}
\begin{figure} % 单栏布局
    \centering
    \includegraphics[width=1.0\linewidth]{ 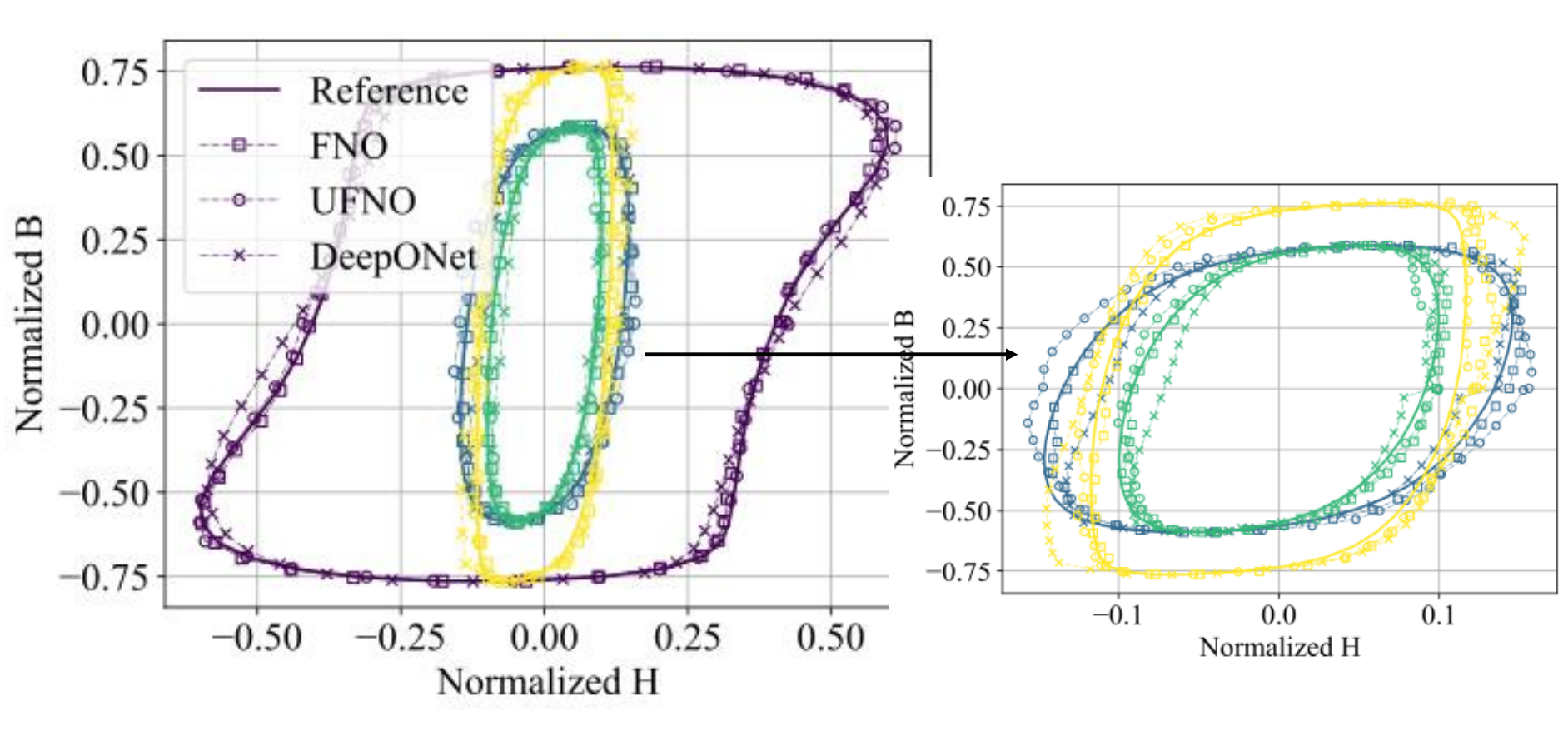}
    \caption{$B-H$ predictions by FNO, U-FNO, and DeepONet, comparing their performance on 4 test loops.}
    \label{fig:preliminary_predictions}
\end{figure}

The input data is structured with two channels: the first channel represents the magnetic flux density $B$ and the second channel represents the time $t$ to distinguish the frequency of each signal. The input data has a size of [36, 2, 500] that corresponds to the 36 DHLs (4 peak values and 9 frequencies), 2 channels and 500 points in per sample.
Considering that the dataset is small, with only 36 DHLs in total, there is no validation data assigned for this case. All data are shuffled and split into training and test data by a ratio of $9\colon1$.\\
As illustrated in section \ref{sec:FNO}, the input data is firstly lifted into higher channels by $\mathcal{P}$ (Fig. \ref{fig:FNO_general_structure}), which can be chosen as any type of NN. We used the simplest type, linear neural network, to lift the channel from 2 into 64, which is followed by 4 blocks of FNO layers. After tuning the hyper-parameters, the width of FNO layers and the selected modes are 64 and 16, respectively. The activation function is set as $ReLU$, the optimizer is $Adam$ with learning rate as $1 \times 10^{-3}$.
% The activation function $g$ is set as $Tanh$. \zg{The learning rate is $1 \times 10^{-3}$.
% The exponential learning rate decay is employed, starting at $1 \times 10^{-3}$ and decreasing by $10 \%$ every 500 steps. This approach enables faster convergence in the early stages and gradual fine-tuning in later stages, ensuing stable optimization \cite{li_exponential_learning_rate_decay_2019}. 
Following the findings in \cite{zongyi_U_FNO_2022} regarding the superior convergence properties of L2-norm loss optimization, we employ the L2-norm between model predictions and reference values as our loss function
% \begin{equation}
%     \label{eq:MSE}
%     \text{MSE} = \frac{1}{n} \sum_{i=1}^{n} (H_i - \hat{H}_i)^2\,,
% \end{equation}
\begin{equation}
    \label{eq:L2_norm}
    L_2\text{-norm} = \sqrt{\sum_{i=1}^n (H_i - \hat{H}_i)^2}
\end{equation}
where $\hat{H_i}$ is the prediction and $H_i$ is the corresponding reference value. After 300 epochs in 39.9\,s, the training loss decreases to $1 \times 10^{-1}$.
% By plotting the predicted $H$ against the corresponding $B$ from the test data, the resulting hysteresis loops are shown in Fig. \ref{fig:FNO_preliminary_results}. 
The predicted hysteresis loops are presented and compared with reference loops as shown in Fig. \ref{fig:preliminary_predictions}.
% The MSE of all test data is $3.76 \times 10^{-4}$. 
% Besides the MSE, also L2 Norm as \eqref{eq:L2_norm} is calculated to verify the accuracy of the trained model, which is $0.87$.\\

Moreover, with the predicted $B - H$ loops as Fig. \ref{fig:preliminary_predictions}, 
the core losses, represented by the area under the $H-B$ loop can be calculated as
% the core losses can be calculated with the integral as \eqref{eq:area_loops}, which equals to the areas of the loops. 
\begin{equation}
\label{eq:area_loops}
P=\frac{1}{T} \int_{B(0)}^{B(T)} H(t) d B(t),
\end{equation}
with period T = $1/f$. Note that, $B$ and $H$ are both denormalized to their respective original values in the numerical integration Eq. \eqref{eq:area_loops}, computed with Simpson's rule. The mean relative error (MRE) between the predication and the reference iron losses is defined as
% \begin{equation}
% \label{eq:RE}
% \text{RE}_i = \left| \frac{P_i - \hat{P}}{P} \right|
% \end{equation}
\begin{equation}
\label{eq:MRE}
\text{MRE} = \frac{1}{n} \sum_{i=1}^{n} \left| \frac{P_i - \hat{P}_i}{P_i} \right|,
\end{equation}
with $P_i$ the reference losses and $\hat{P_i}$ the predicted losses for sample $i$. The calculated MREs of all test data are listed in Table. \ref{tab:model_comparison}, labeled as ``No'' (no augmentation).
\subsubsection{U-FNO}
The training dataset is kept the same as in the previous case. Similar to Fig. \ref{fig:FNO_general_structure}, a linear NN firstly lifts the input from 2 channels to 160, which is followed by 3 CNNs (CNN1, CNN2 and CNN3 in Fig. \ref{fig:U_net_structure}) with 160 input and output channels of kernel size 3.
As a result, the number of channels for \textit{$Out\_conv1$}, \textit{$Out\_conv2$} and \textit{$Out\_conv3$} is 160. In contrast, for the de-convolution NNs, De-CNN2 has 160 channels for input and output, while De-CNN1 and De-CNN0 have 360 and 160 channels for input and output, respectively, as their inputs are formed by the concatenation of \textit{$Out\_conv$} and \textit{$Out\_deconv$}. 
The suitable kernel size, stride and padding for each CNN are chosen to make the outputs of each model part match. Finally, after all U-FNO layers, the linear NN is applied to decrease the channel number from 160 to the output channel number 1. The FFT modes, loss function, activation function and learning rate are kept the same as in FNO.
Note that we use the same training data as FNO, after 300 epochs for 412.2\,s, the loss decreases to $1 \times 10^{-1}$. The predicted hysteresis loops are plotted in Fig. \ref{fig:preliminary_predictions}
% , with MSE \zg{of} $1.18 \times 10^{-4}$. 
The calculated core losses and MRE of test data are listed in Table. \ref{tab:model_comparison} labeled as ``No'' (no augmentation).
\subsubsection{DeepONet}
For frequency information of data, in DeepONet, instead of adding another channel with corresponding $t$, we appended $f$ to the end of the input $B$. So, the input shape of the branch net is $[36,501]$ and the shape of output $H$ is $[36,500]$. 
Time $t$ for each frequency is normalized by the period, $T = 1/f$, so $t$ is in range [0,1] for each signal, which is the input for trunk net.
% The trunk net takes the time $t$ as input, where $t$ is fixed to the interval [0,1] with 500 equidistant points for all signals, regardless of their frequencies. 
This ensures a consistent time scale across all dataset, while the frequency information is incorporated separately into the branch net.
Considering that the dataset is small and the structure is simple, the branch net and trunk net are both chosen as the multilayer perceptron (MLP), with the input size as 501 and 1 while 
have the same depth, neurons in each layer and also output size as 8, 100 and 100 respectively.
% have the same depth as 8, and neurons in each layer as 250 and the same output size as 100. 
% Then, the outputs of 2 nets multiply and sum to generate the final output. 
The outputs of the branch net and the trunk net are combined via element-wise multiplication and summation to generate the final output.
The activation function, loss function, optimizer and the learning rate are the same as for FNO and U-FNO.
% The loss function of the model for training is MSE, the optimizer and the activation function are chosen as $Adam$ and $ReLU$, respectively, with the same learning rate. The training and test data are the same as for FNO and U-FNO. 
After 6000 epoch with 76.4\,s, the loss decreases to $1 \times 10^{-1}$. The hysteresis loops and corresponding reference values are in Fig. \ref{fig:preliminary_predictions}. The MRE of calculated iron losses are listed in Table. \ref{tab:model_comparison}.
% Fig. \ref{fig:Bar_plot_without_rolling} shows the MSE and mean relative error (MRE) of all test data calculated by \eqref{eq:MRE} for each model \zg{with $n$ as the number of test data}.
In terms of computational efficiency and accuracy, the FNO model demonstrates superior performance, achieving the minimal MRE with the shortest training duration among all tested models.
\subsection{Optimized results with data processing}
To improve the generalization and robustness of models, data augmentation becomes a necessary step, especially when dealing with limited or highly structured datasets \cite{Shorten_data_augmentation_2019}. By introducing controlled variations in the input, augmentation helps the network learn invariant features and reduces overfitting. In this work, two augmentation strategies are applied while keeping the network architectures identical to the previous.
\subsubsection{Cyclic rolling augmentation}
\begin{figure} % 单栏布局
    \centering
    \includegraphics[width=0.8\linewidth]{ 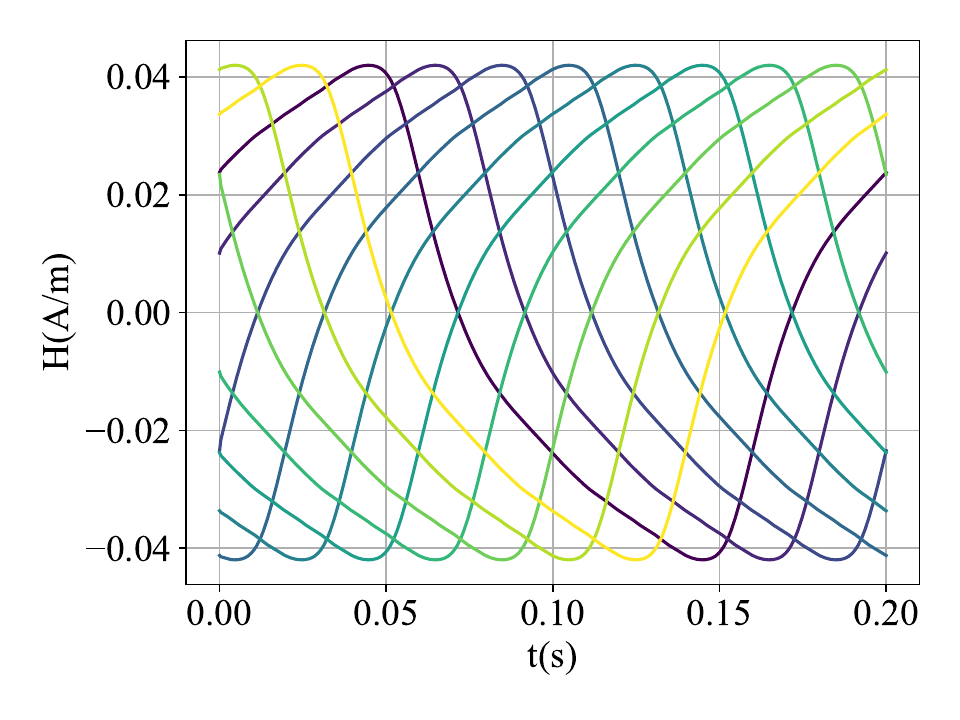}
    \caption{Resulting $H$ after rolling normalized $H$ with $B_{peak} = 1.0$T and $f = 5$Hz by 10 equidistant phase shift as $0.2\pi$.}
    \label{fig:Rolling_H}
\end{figure}

In real-world scenarios, applied signals often exhibit phase shifts. To evaluate the ability of the trained model to predict the corresponding $H$ of $B$ with phase shifts, a sinusoidal $B$ with phase shift of $0.2\pi$ was used to test the preliminary trained FNO, U-FNO and DeepONet models. However, all three models failed to predict the corresponding $H$. This limitation arises because NNs treat datasets with operations such as cyclic rolling as entirely new data \cite{Magnet_Haoran_2023}, in other words, the periodicity information is lost.
% When the phase shifts vary, the starting points of the hysteresis loops also differ. While experimental or numerical methods primarily adjust the initial values to account for phase shifts, NNs perceive each phase shift as a distinct dataset. 
As a result, a model trained only on signals without shifts struggles to generalize and predict hysteresis for signals with any phase shifts. To address this issue, cyclic rolling augmentation strategy, as illustrate in Fig. \ref{fig:Rolling_H}, provides an effective solution. 
Cyclic rolling consists in shifting elements in an array cyclically, i.e. elements shifted out from one end, reenter at the other end.
By rolling the data, datasets with various phase shifts can be generated, enabling the model to be trained on a more diverse dataset. In addition, this approach not only enhances the ability of models to handle phase shifts but also serves as a method for augmenting the training data \cite{Magnet_Haoran_2023}. In this study, each data pair from the 36 $B - H$ loops are subjected to an equidistant phase shift as $0.2\pi$, resulting in 10 corresponding data sets for each, as 360 in total. The total dataset are split into the training, validation and test data by the ratio as $8\colon1\colon1$.\\
\begin{figure} % 单栏布局
    \centering
    \includegraphics[width=1\linewidth]{ 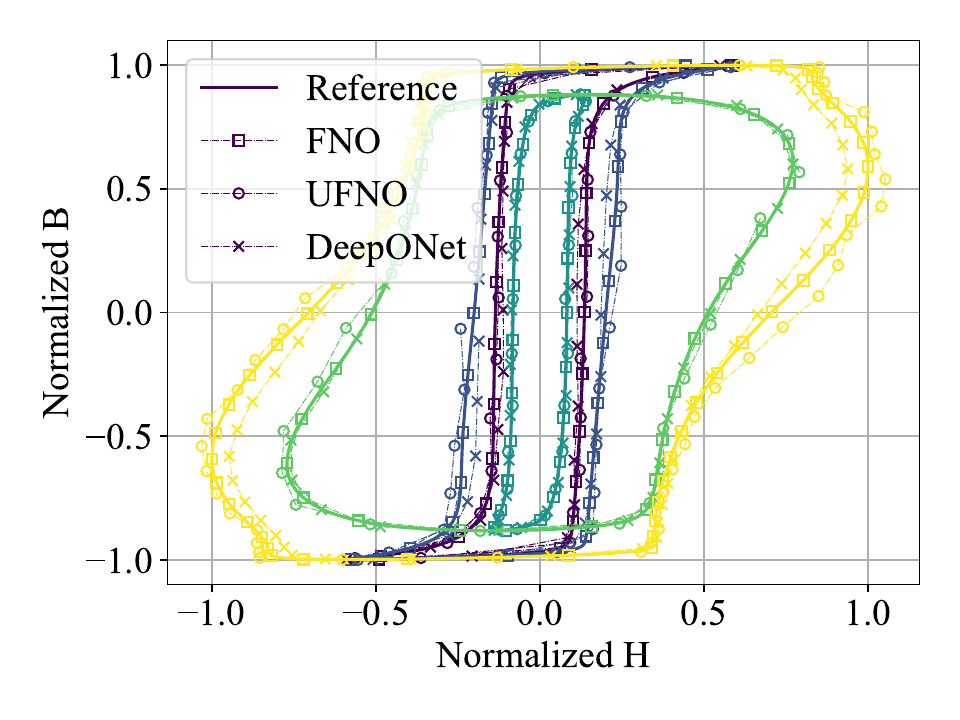}
    \caption{$B-H$ predictions by FNO, U-FNO, and DeepONet with training data after cyclic rolling, comparing their performance on 5 test loops.}
    \label{fig:rolling_predictions}
\end{figure}
For visualizing the predictions, five loops of test data ranging different frequencies and peak values are shown in Fig. \ref{fig:rolling_predictions}. In addition, the MRE of the calculated iron losses between the prediction and reference are listed in Table. \ref{tab:model_comparison} labeled as ``Cyclic''. The small errors indicate that all three models trained on the data augmented through cyclic rolling effectively handle the phase shift information, with FNO showing the best performance.
%
% %%%%%%%%%%%%%%%%%%%%%%%
\begin{figure} % 单栏布局
    \centering
    \includegraphics[width=1\linewidth]{ 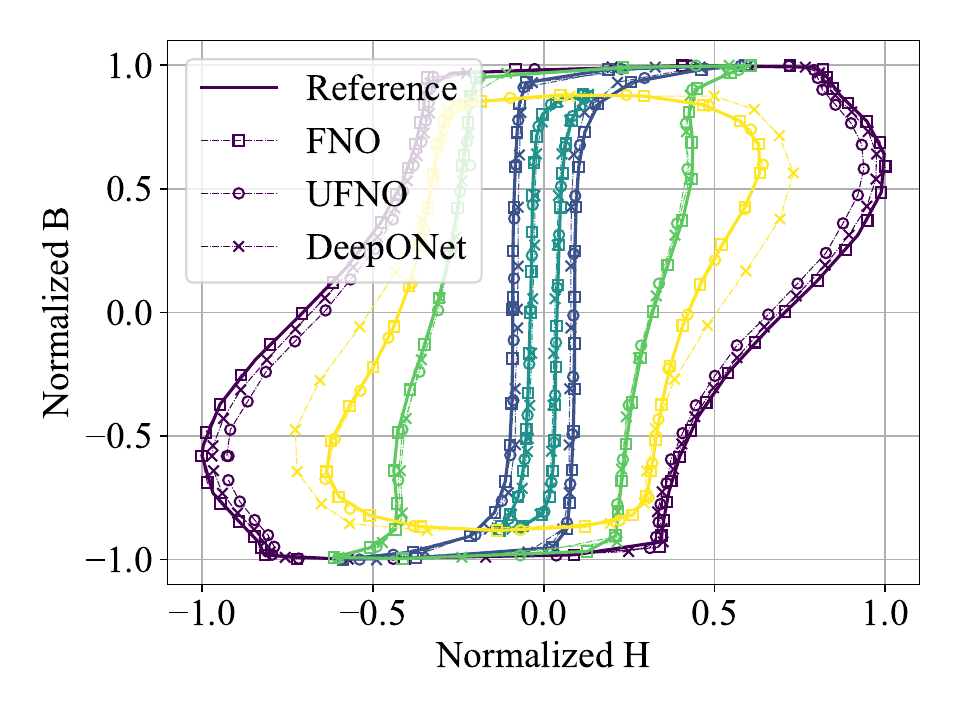}
    \caption{$B-H$ predictions by FNO, U-FNO, and DeepONet with training data after cyclic rolling and GDA, comparing their performance on 5 test loops.}
    \label{fig:GDA_predictions}
\end{figure}
%%%%%%%%%%%%%%%%%%%%%%%
\subsubsection{Gaussian data augmentation (GDA)}
In the context of NNs, GDA is a commonly used technique to introduce Gaussian noise into training data \cite{nawa_prediction_accuracy_improvement_gaussian_2023}. This approach enables the NNs to explore new samples that closely resemble the original data, thereby increasing dataset diversity, smoothing the input structure, and improving data accessibility for learning \cite{zhang_LIB_gaussian_2024}. In order to further enhance the prediction accuracy, in this study, GDA is implemented.
The process is conducted as follows:
we consider the $360$ groups of data obtained after cyclic rolling as the original training dataset.
To introduce variability, the random noise following the Gaussian distribution
\begin{equation}
\label{eq:gaussian_function}
    f(x) = \frac{1}{\sqrt{2\pi\sigma^2}} \exp\left( -\frac{(x - \mu)^2}{2\sigma^2} \right)\,,
\end{equation}
with mean ($\mu$ = 0) and the standard deviation ($\sigma$ = 0.05) is generated. These Gaussian noises are independently added to each original input sample $B$, while maintaining the corresponding original output values, to generate additional training samples. The artificially generated data is then concatenated with the original data for training and improving the robustness, generalization capabilities of models.\\
Then, the size of data are now [720,500], with half the original data and half with added Gaussian noise. The corresponding MRE (Eq.\eqref{eq:MRE}) of the calculated iron losses for each model configuration is shown in Table. \ref{tab:model_comparison} 
% Equation \eqref{eq:MRE} is used to calculate the corresponding MRE of the calculated core losses for each configuration of models shown in Table. \ref{tab:model_comparison}
labeled as ``Cyclic + GDA''. 
Five loops of test data with different frequencies and peak values are selected and shown in Fig. \ref{fig:GDA_predictions}.
To quantify the performance enhancement achieved through data augmentation methods, the improvement metric $\eta$ is defined as
 \begin{equation}
     \label{eq:improvement}
     \text{$\eta$ (\%)} = \left( 1 - \frac{\text{MRE (With augmentation)}}{\text{MRE (No augmentation)}} \right)\times 100,
 \end{equation}
$\eta$ by two augmentation methods are summarized in Table. \ref{tab:model_comparison}.
We observe that both data augmentation methods significantly increase the accuracy of all models. In every case, FNO consistently achieved the lowest error.
\begin{table}[h!]
\centering
\caption{Summary of model performances}
\begin{tabular}{lcccc}
\toprule
Model & Augmentation & MRE & $\eta$($\%$) \\
\midrule
FNO & No &  2.27 $\times 10 ^{-2}$   & &  \\
 & Cyclic & $0.97 \times 10^{-3}$ & 95.74\\
 & Cyclic + GDA & $0.24 \times 10^{-3}$ & 98.95 \\
\cmidrule{1-4}
U-FNO & No & 4.90 $\times 10 ^{-2}$ & \\
 & Cyclic & $3.80 \times 10^{-2}$ & 22.53\\
 & Cyclic + GDA & $2.55 \times 10^{-2}$ & 47.96 \\
\cmidrule{1-4}
DeepONet & No & 9.60 $\times 10 ^{-2}$  & \\
 & Cyclic & $6.81 \times 10^{-2}$ & 29.06\\
 & Cyclic + GDA & $4.29 \times 10^{-2}$ & 55.34 \\
\bottomrule
\end{tabular}
\label{tab:model_comparison}
\end{table}
\section{Conclusion}
In this study, a novel dynamic hysteresis model for GO ferromagnetic materials based on neural operators is proposed. Three types of neural operators including FNO, U-FNO and also DeepONet are investigated.
To enhance the diversity of dataset and increase the accuracy of the models, data augmentation strategies including cyclic rolling augmentation and Gaussian data augmentation are applied to the experimental data.
Additionally, iron losses are calculated and compared based on the predicted $B - H$ curves. 
The performance of the models is evaluated by comparing the MRE of iron losses between the predictions and the reference values. 
The results demonstrate that all three models can effectively capture the hysteresis behavior of the GO material, considering not only the peak values but also frequency and phase shift effects.
Moreover, with the data augmentation strategies, the accuracy of each model is further enhanced.
Among the evaluated models, FNO consistently outperforms the others across different scenarios, exhibiting superior accuracy and generalization capabilities. 
These findings highlight the potential of neural operator-based approaches for accurate and efficient dynamic hysteresis modeling of GO steels.
\FloatBarrier
\section*{Acknowledgment}
Ziqing Guo acknowledges support from China Scholarship Council (CSC), No. 202206280041, in the preparation of this manuscript.
\ifCLASSOPTIONcaptionsoff
  \newpage
\fi

% \begin{IEEEbiography}{Ziqing Guo}
% Biography text here.
% \end{IEEEbiography}
\bibliographystyle{ieeetr}        % 或你使用的其他样式
\bibliography{ bare_jrnl_transmag} % 不加 .bbl 后缀

\begin{thebibliography}{10}

\bibitem{Ferr_core_loss_Roshen_1991}
W.~Roshen, ``Ferrite core loss for power magnetic components design,'' {\em IEEE Transactions on Magnetics}, vol.~27, no.~6, pp.~4407--4415, 1991.

\bibitem{chen_dynamic_hysteresis_model_2022}
L.~Chen, Z.~Zhang, T.~Ben, and H.~Zhao, ``Dynamic magnetic hysteresis modeling based on improved parametric magneto-dynamic model,'' {\em IEEE Transactions on Applied Superconductivity}, vol.~32, no.~6, pp.~1--5, 2022.

\bibitem{zirka_dynamic_JA_2015}
S.~Zirka, Y.~Moroz, S.~Steentjes, K.~Hameyer, K.~Chwastek, S.~Zurek, and R.~Harrison, ``Dynamic magnetization models for soft ferromagnetic materials with coarse and fine domain structures,'' {\em Journal of Magnetism and Magnetic Materials}, vol.~394, pp.~229--236, 2015.

\bibitem{Florent_liege_NN_2024}
F.~Purnode, F.~Henrotte, G.~Louppe, and C.~Geuzaine, ``Neural network-based simulation of fields and losses in electrical machines with ferromagnetic laminated cores,'' {\em International Journal of Numerical Modelling: Electronic Networks, Devices and Fields}, vol.~37, no.~2, p.~e3226, 2024.

\bibitem{cardelli_comparison_hysteresis_models_2023}
E.~Cardelli, A.~Faba, A.~Laudani, S.~Quondam~Antonio, and A.~M. Ghanim, ``Comparison between different models of magnetic hysteresis in the solution of the {TEAM} 32 problem,'' {\em International Journal of Numerical Modelling: Electronic Networks, Devices and Fields}, vol.~36, no.~6, p.~e3103, 2023.

\bibitem{corti_dynamic_core_losses_2020}
F.~Corti, A.~Reatti, E.~Cardeli, A.~Faba, and H.~Rimal, ``Improved spice simulation of dynamic core losses for ferrites with nonuniform field and its experimental validation,'' {\em IEEE Transactions on Industrial Electronics}, vol.~68, no.~12, pp.~12069--12078, 2020.

\bibitem{jiles1986theory}
D.~C. Jiles and D.~L. Atherton, ``Theory of ferromagnetic hysteresis,'' {\em Journal of magnetism and magnetic materials}, vol.~61, no.~1-2, pp.~48--60, 1986.

\bibitem{Mayergoyz_dynamic_preisach_1988}
I.~Mayergoyz, ``Dynamic preisach models of hysteresis,'' {\em IEEE Transactions on Magnetics}, vol.~24, no.~6, pp.~2925--2927, 1988.

\bibitem{zirka_static_dynamic_hysteresis_model_2011}
S.~E. Zirka, Y.~I. Moroz, A.~J. Moses, and C.~M. Arturi, ``Static and dynamic hysteresis models for studying transformer transients,'' {\em IEEE Transactions on Power Delivery}, vol.~26, no.~4, pp.~2352--2362, 2011.

\bibitem{Bertotti_loss_separation_1988}
G.~Bertotti, ``General properties of power losses in soft ferromagnetic materials,'' {\em IEEE Transactions on Magnetics}, vol.~24, no.~1, pp.~621--630, 1988.

\bibitem{hamed_GO_model_2021}
H.~Hamzehbahmani, ``Static hysteresis modeling for grain-oriented electrical steels based on the phenomenological concepts of energy loss mechanism,'' {\em Journal of Applied Physics}, vol.~130, no.~5, 2021.

\bibitem{du_dynamic_JA_losses_high_freq_2014}
R.~Du and P.~Robertson, ``Dynamic {Jiles--Atherton} model for determining the magnetic power loss at high frequency in permanent magnet machines,'' {\em IEEE Transactions on Magnetics}, vol.~51, no.~6, pp.~1--10, 2014.

\bibitem{tian_backpropogation_NN_2021}
M.~Tian, H.~Li, and H.~Zhang, ``Neural network model for magnetization characteristics of ferromagnetic materials,'' {\em IEEE Access}, vol.~9, pp.~71236--71243, 2021.

\bibitem{grech_RNN_Preisach_2020}
C.~Grech, M.~Buzio, M.~Pentella, and N.~Sammut, ``Dynamic ferromagnetic hysteresis modelling using a {Preisach-recurrent} neural network model,'' {\em Materials}, vol.~13, no.~11, p.~2561, 2020.

\bibitem{ding_LSTM_CNN_2024}
C.~Ding, Y.~Bai, Y.~Ji, and P.~Ma, ``Neural network modeling of complex hysteresis loops in ferromagnetic materials,'' {\em IEEJ Transactions on Electrical and Electronic Engineering}, 2024.

\bibitem{Magnet_Haoran_2023}
H.~Li, D.~Serrano, T.~Guillod, S.~Wang, E.~Dogariu, A.~Nadler, M.~Luo, V.~Bansal, N.~K. Jha, Y.~Chen, C.~R. Sullivan, and M.~Chen, ``How {MagNet}: Machine learning framework for modeling power magnetic material characteristics,'' {\em IEEE Transactions on Power Electronics}, vol.~38, no.~12, pp.~15829--15853, 2023.

\bibitem{chen_universal_app_theory_1995}
T.~Chen and H.~Chen, ``Universal approximation to nonlinear operators by neural networks with arbitrary activation functions and its application to dynamical systems,'' {\em IEEE transactions on neural networks}, vol.~6, no.~4, pp.~911--917, 1995.

\bibitem{lulu_Deeponet_2021}
L.~Lu, P.~Jin, G.~Pang, Z.~Zhang, and G.~E. Karniadakis, ``Learning nonlinear operators via {DeepONet} based on the universal approximation theorem of operators,'' {\em Nature machine intelligence}, vol.~3, no.~3, pp.~218--229, 2021.

\bibitem{lulu_deeponet_2019}
L.~Lu, P.~Jin, and G.~E. Karniadakis, ``{DeepONet}: Learning nonlinear operators for identifying differential equations based on the universal approximation theorem of operators,'' {\em arXiv preprint arXiv:1910.03193}, 2019.

\bibitem{zongyi_FNO_2020}
Z.~Li, N.~Kovachki, K.~Azizzadenesheli, B.~Liu, K.~Bhattacharya, A.~Stuart, and A.~Anandkumar, ``Fourier neural operator for parametric partial differential equations,'' {\em arXiv preprint arXiv:2010.08895}, 2020.

\bibitem{Lyu_fluid_flow_NO_2023}
Y.~Lyu, X.~Zhao, Z.~Gong, X.~Kang, and W.~Yao, ``Multi-fidelity prediction of fluid flow based on transfer learning using fourier neural operator,'' {\em Physics of Fluids}, vol.~35, no.~7, 2023.

\bibitem{TU_Delft_NO_2024}
A.~Chandra, T.~Kapoor, M.~Curti, K.~Tiels, and E.~A. Lomonova, ``Characterizing nonlinear piezoelectric dynamics through deep neural operator learning,'' {\em Applied Physics Letters}, vol.~125, no.~26, 2024.

\bibitem{yuan_heat_transfer_NO_2025}
J.~Yuan, L.~Zeng, and Y.~Gui, ``Method for predicting conductive heat transfer topologies based on fourier neural operator,'' {\em International Communications in Heat and Mass Transfer}, vol.~160, p.~108332, 2025.

\bibitem{chandra_no_magnetic_2024}
A.~Chandra, B.~Daniels, M.~Curti, K.~Tiels, and E.~A. Lomonova, ``Magnetic hysteresis modeling with neural operators,'' {\em IEEE Transactions on Magnetics}, 2024.

\bibitem{Marion_JA_model_parameters_2008}
R.~Marion, R.~Scorretti, N.~Siauve, M.-A. Raulet, and L.~Krahenbuhl, ``Identification of {Jiles–Atherton} model parameters using particle swarm optimization,'' {\em IEEE Transactions on Magnetics}, vol.~44, no.~6, pp.~894--897, 2008.

\bibitem{zguo_PINN_2025}
Z.~Guo, B.~Nguyen, and R.~V. Sabariego, ``Physics-informed neural network for solving 1d nonlinear time-domain magneto-quasi-static problems,'' {\em IEEE Transactions on Magnetics}, 2025.

\bibitem{kovachki_zongyi_FNO_2023}
N.~Kovachki, Z.~Li, B.~Liu, K.~Azizzadenesheli, K.~Bhattacharya, A.~Stuart, and A.~Anandkumar, ``Neural operator: Learning maps between function spaces with applications to pdes,'' {\em Journal of Machine Learning Research}, vol.~24, no.~89, pp.~1--97, 2023.

\bibitem{kaiming_residual_connection_2016}
K.~He, X.~Zhang, S.~Ren, and J.~Sun, ``Deep residual learning for image recognition,'' in {\em Proceedings of the IEEE conference on computer vision and pattern recognition}, pp.~770--778, 2016.

\bibitem{siddique_U_NET_2021}
N.~Siddique, S.~Paheding, C.~P. Elkin, and V.~Devabhaktuni, ``{U-net} and its variants for medical image segmentation: A review of theory and applications,'' {\em IEEE access}, vol.~9, pp.~82031--82057, 2021.

\bibitem{zongyi_U_FNO_2022}
G.~Wen, Z.~Li, K.~Azizzadenesheli, A.~Anandkumar, and S.~M. Benson, ``{U-FNO—An} enhanced fourier neural operator-based deep-learning model for multiphase flow,'' {\em Advances in Water Resources}, vol.~163, p.~104180, 2022.

\bibitem{Shorten_data_augmentation_2019}
C.~Shorten and T.~M. Khoshgoftaar, ``A survey on image data augmentation for deep learning,'' {\em Journal of big data}, vol.~6, no.~1, pp.~1--48, 2019.

\bibitem{nawa_prediction_accuracy_improvement_gaussian_2023}
K.~Nawa, K.~Hagiwara, and K.~Nakamura, ``Prediction-accuracy improvement of neural network to ferromagnetic multilayers by {Gaussian} data augmentation and ensemble learning,'' {\em Computational Materials Science}, vol.~219, p.~112032, 2023.

\bibitem{zhang_LIB_gaussian_2024}
C.~Zhang, Y.~Zhang, Z.~Li, Z.~Zhang, M.~S. Nazir, and T.~Peng, ``Enhancing state of charge and state of energy estimation in lithium-ion batteries based on a {TimesNet} model with {Gaussian} data augmentation and error correction,'' {\em Applied Energy}, vol.~359, p.~122669, 2024.

\end{thebibliography}
\end{document}